\documentclass{PoS}

\usepackage{nicefrac}
\usepackage{subfig}

\newcommand\T{\rule{0pt}{2.6ex}}

\title{$B$ and $B_s$ semileptonic decay form factors with NRQCD/HISQ quarks}

\ShortTitle{$B$ and $B_s$ semileptonic decay form factors with NRQCD/HISQ quarks}

\author{
C.~M.~Bouchard$^{\, *\, a}$,
G.~Peter~Lepage$^{\, b}$,
Chris~J.~Monahan$^{\, c}$,
Heechang~Na$^{\, d, e}$,
Junko~Shigemitsu$^{\, a}$
\hphantom{\speaker{C.M.~Bouchard}}
\\ \\
\llap{$^a$}Department of Physics, The Ohio State University, Columbus, OH 43210, USA\\
\llap{$^b$}Laboratory of Elementary Particle Physics, Cornell University, Ithaca, NY 14853, USA\\
\llap{$^c$}Department of Physics, College of William and Mary, VA 23187-8795, USA\\
\llap{$^d$}Argonne Leadership Computing Facility, Argonne National Laboratory, Argonne, IL 60493, USA\\
\llap{$^e$}Physics Department, University of Utah, Salt Lake City, UT 84112, USA}
\author{HPQCD Collaboration\\
\email{bouchard.18@osu.edu}}

\abstract{
We discuss our ongoing effort to calculate form factors for several $B$ and $B_s$ semileptonic decays. We have recently completed the first unquenched calculation of the form factors for the rare decay $B\to K \ell \ell$. Extrapolated over the full kinematic range of $q^2$ via model-independent $z$ expansion, these form factor results allow us to calculate several Standard Model observables. We compare with experiment (Belle, BABAR, CDF, and LHCb) where possible and make predictions elsewhere. We discuss preliminary results for $B_s\to K\ell\nu$ which, when combined with anticipated experimental results, will provide an alternative exclusive determination of $|V_{ub}|$. We are exploring the possibility of using ratios of form factors for this decay with those for the unphysical decay $B_s\to\eta_s$ as a means of significantly reducing form factor errors. We are also studying $B\to\pi\ell\nu$, form factors for which are combined with experiment in the standard exclusive determination of $|V_{ub}|$. Our simulations use NRQCD heavy and HISQ light valence quarks on the MILC $2+1$ dynamical asqtad configurations.
}

\FullConference{31st International Symposium on Lattice Field Theory, LATTICE 2013\\
                 July 29 -- August 3, 2013\\
                 Mainz, Germany}

\begin{document}

\section{$B \to K \ell \ell$}

The $b\to s$ FCNC associated with this decay probes new physics.  Such processes have recently received significant theoretical~\cite{theory} and experimental~\cite{expt} attention with future scrutiny planned at LHCb and BelleII.
In~\cite{PRD} we reported on the first unquenched lattice calculation of the form factors for $B\to K\ell\ell$ and in~\cite{PRL} explored the Standard Model (SM) implications.  This section outlines the approach and results of these works. 
Other unquenched lattice works are currently underway~\cite{Liu, ASKtalk}.

\subsection{Setup and simulation}
\label{sec-setup}

The SM $V-A$ interaction leads to hadronic vector matrix elements.  In the $B$ rest frame
\begin{eqnarray}
\langle K | V^0 | B\rangle &=& \sqrt{2M_B}\ f_\parallel(q^2), \\
\langle K | V^i | B\rangle &=& \sqrt{2M_B}\  p^i_K\ f_\perp(q^2),
\end{eqnarray}
where $q^\mu=p^\mu_B - p^\mu_K$ is the four-momentum transferred to the leptons.
The phenomenologically relevant form factors $f_{0,+}$ are related to $f_{\parallel, \perp}$ by
\begin{eqnarray}
f_0 &=& \frac{\sqrt{2M_B}}{M_B^2-M_K^2} \left[ (M_B-E_K) f_\parallel + {\rm \bf p}_K^2\, f_\perp   \right], \\
f_+ &=& \frac{1}{\sqrt{2M_B}}  \left[  f_\parallel + (M_B - E_K) f_\perp  \right].
\end{eqnarray}
Generic new physics is characterized by the addition of a tensor current and associated form factor
\begin{equation}
\langle K | T^{j0} | B \rangle = \frac{2i\, M_B p^j_K}{M_B+M_K}\ f_T(q^2).
\end{equation}
%
%
\begin{table*}[t]
\begin{tabular}{ccccccccc}
\hline\hline	
	\T ens   	& $L^3\times N_t$	& $\approx a$ [fm]		& $m_l^{\rm sea}/m_s^{\rm sea}$	&  $N_{\rm conf}$	& $N_{\rm tsrc}$	& $am_l^{\rm val}$	& $am_s^{\rm val}$	& $T$		\\ [0.5ex]
	\hline
	\T C1 	& $24^3 \times 64$	& 0.12				& 0.005/0.050					& 1200			& 2				& 0.007			& 0.0489			& 13, 14, 15	\\
	\T C2	& $20^3 \times 64$	& 0.12				& 0.010/0.050					& 1200			& 2				& 0.0123			& 0.0492			&13, 14, 15 	\\
	\T C3	& $20^3 \times 64$	& 0.12 				& 0.020/0.050					& 600			& 2				& 0.0246			& 0.0491			& 13, 14, 15	\\ 
	\T F1		& $28^3 \times 96$	& 0.09 				& 0.0062/0.031					& 1200			& 4				& 0.00674			& 0.0337			& 23, 24		\\
	\T F2		& $28^3 \times 96$	& 0.09 				& 0.0124/0.031					& 600			& 4				& 0.0135			& 0.0336			& 21, 22, 24	\\ [0.5ex]
\hline\hline
\end{tabular}\caption{Left to right:  ensemble, lattice volume, lattice spacing, light/strange sea-quark masses, number of configurations, number of time sources, valence light-quark mass, valence strange-quark mass, and temporal separations between the $B$ meson and the kaon.}
\label{tab-ens}
\end{table*}
We simulate with the MILC $2+1$ asqtad gauge configurations~\cite{MILC} shown in Table~\ref{tab-ens}.  We use valence NRQCD~\cite{NRQCD} $b$ quarks, tuned in~\cite{btune}, and HISQ~\cite{HISQ} light and strange quarks, propagators for which were generated for~\cite{lprop,sprop}.

\subsection{Correlator fits and matching}
\label{sec-fits}

Values of lattice matrix elements are extracted from two and three point correlation function data via Bayesian fits~\cite{bayes}.
Correlator data in the fits includes local and smeared $B$ mesons, four momenta $\nicefrac{ L{\rm \bf p}_K}{2\pi}  \in \{ 000, 100, 110, 111 \}$, three currents $V_0$, $V_k$, and $T_{\mu\nu}$, and the $B$ meson and kaon temporal separations of Table~\ref{tab-ens}.
Two and three point data are fit simultaneously and correlations among form factors at different momenta and for each current are determined.
Two point fit parameters are verified to be consistent between simultaneous fits and fits to just the two point data.
Kaon fit results are found to satisfy the dispersion relation to $\mathcal{O}(\alpha^2_s a^2)$.
Lattice values of the matrix elements are matched to the continuum $\overline{\rm MS}$ scheme through $\mathcal{O}(\alpha_s, 1/(am_b), \Lambda_{\rm QCD}/m_b)$ using massless HISQ, one-loop lattice perturbation theory~\cite{match}.

\subsection{Chiral-continuum and kinematic extrapolations}
\label{sec-extrap}
\begin{figure}[t!]
\hspace{-0.2in}
{\scalebox{1}{\includegraphics[angle=-90,width=0.53\textwidth]{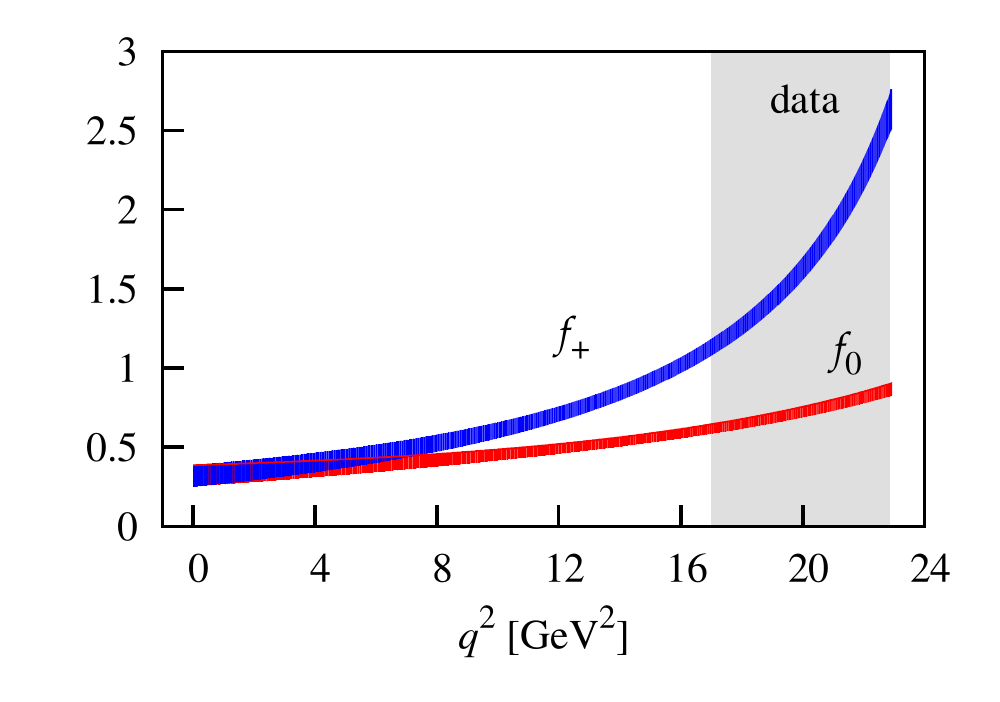}}}
\hspace{-0.2in}
{\scalebox{1}{\includegraphics[angle=-90,width=0.53\textwidth]{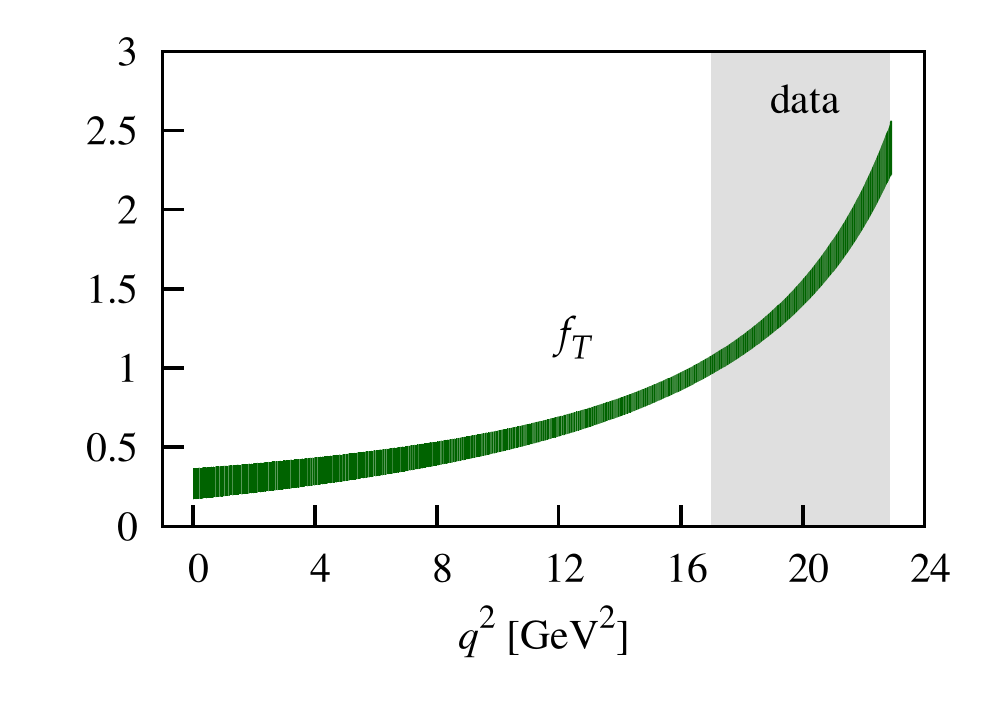}}}
\caption{Form factors for $B\to K\ell \ell$.}
\label{fig-finalFFs}
\end{figure}

The extrapolation to physical light quark mass is performed using NLO chiral logs for $f_{\parallel,\perp}$ from PQ$\chi$PT~\cite{ChPT}.  We account for discretization (and taste breaking) effects via generic $\mathcal{O}(a^2)$ and $\mathcal{O}(a^4)$ mass independent, light quark mass-dependent, and heavy quark mass-dependent discretization terms and include NLO and NNLO chiral analytic terms.
The fit ans\"atze for $f_{0,+}$ are written in terms of $f_{\parallel,\perp}$ and the ansatz for $f_T$ is based on $f_\perp$.  



Results of the chiral/continuum extrapolation are used to generate data points and a covariance matrix for subsequent extrapolation over the full kinematic range of $q^2$.  This extrapolation is performed using the model-independent $z$ expansion~\cite{zexp}
with BCL parameterization~\cite{BCL}
\begin{eqnarray}
f_0(q^2) &=& \sum_{k=0}^K a_k^0\ z(q^2)^k, \\
f_i(q^2) &=& \frac{1}{P_i(q^2)} \sum_{k=0}^{K-1} a_k^i \left[ z(q^2)^k - (-1)^{k-K} \frac{k}{K}\ z(q^2)^K  \right],
\end{eqnarray}
where $i=+,T$.
Poles are removed by the Blaschke factor $P_i(q^2) = 1-q^2/(M_i^{\rm pole})^2$.
%
%
Results of the two-step extrapolation are consistent with those obtained using the modified $z$ expansion~\cite{lprop, sprop}.
Final form factor results, including an additional 4\% systematic error due primarily to matching, are plotted in Fig.~\ref{fig-finalFFs}.
Ref.~\cite{PRD} provides a graphical error budget using methods outlined in~\cite{errmeth}.

\subsection{Phenomenology}

\begin{figure*}[t!]
\hspace{-0.07in}
{\scalebox{0.98}{\includegraphics[angle=-90,width=0.53\textwidth]{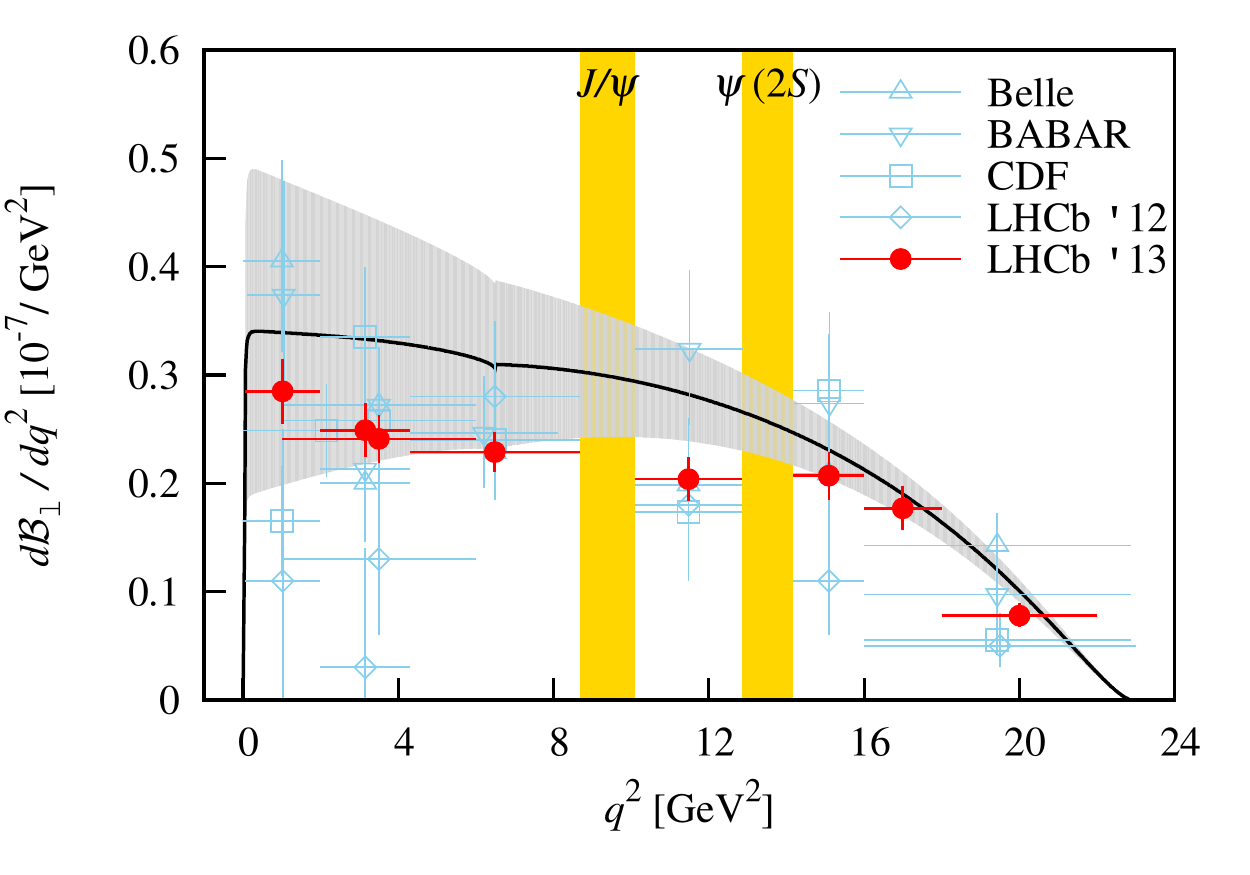}}}
{\scalebox{0.98}{\includegraphics[angle=-90,width=0.53\textwidth]{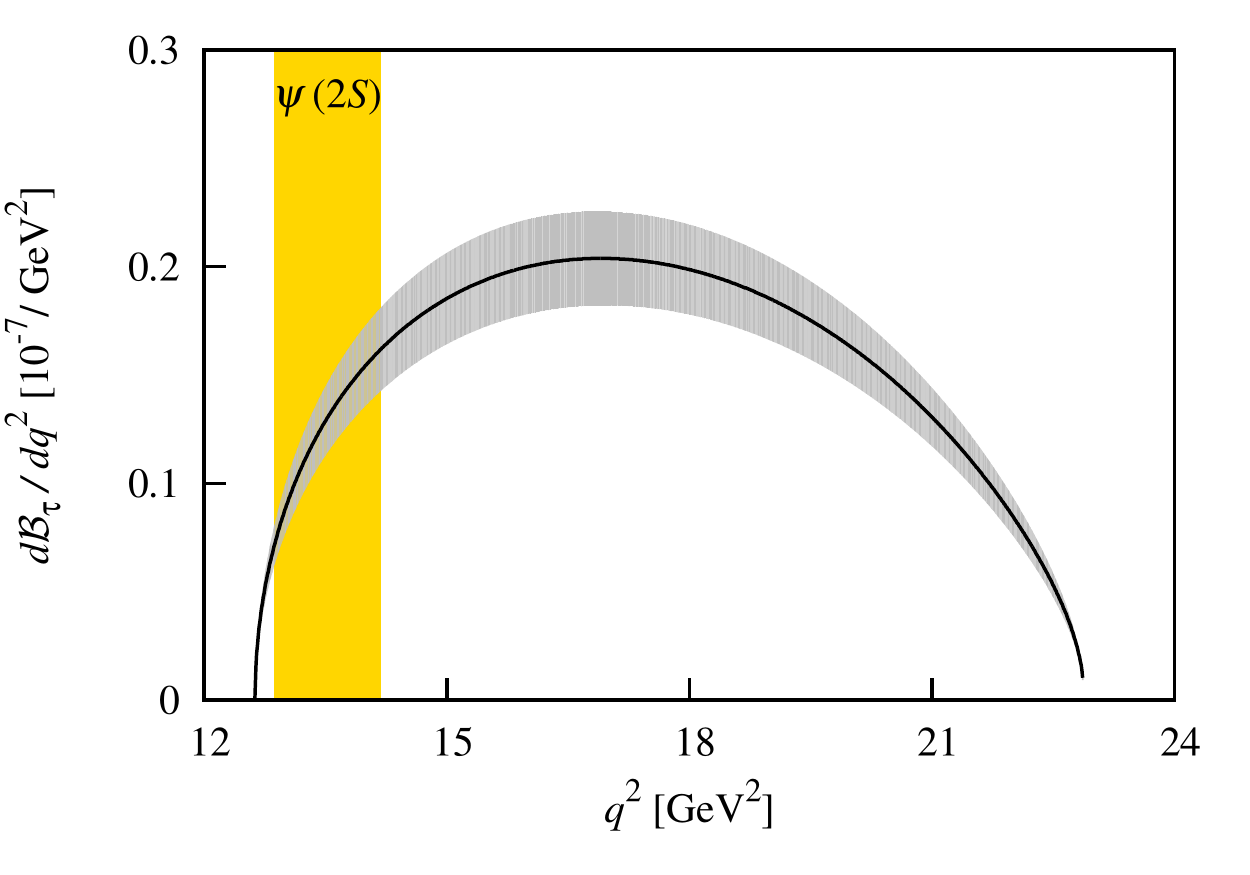}}}
\caption{Standard Model differential branching fraction predictions and experiment for $B\to K\ell\ell$.}
\label{fig-dBdq2}
\end{figure*}

Using the form factor results, Wilson coefficients, and other input parameters, we calculate in~\cite{PRL} the SM differential branching fraction $d \mathcal{B}_\ell (B\to K\ell\ell\,) / dq^2$ over the full range of $q^2$ and in experimentally motivated $q^2$ bins.
Fig.~\ref{fig-dBdq2} compares our results to experiment for light dileptons ($\ell = e, \mu$) and shows our prediction for the ditau final state.
We note the observation of a new $c \bar c$ resonance~\cite{ccbar}, believed to be the $\psi(4160)$, will likely modify the future choice of $q^2$ bins at small recoil (large $q^2$).  Ref.~\cite{PRL} also calculates ratios of branching fractions and the ``flat term" in the angular distribution of the differential decay rate and compares to experiment and other calculations where available.

\section{Preliminary results for other decays}

The setup and simulation outlined in Sec.~\ref{sec-setup} and analysis steps of Secs.~\ref{sec-fits} and~\ref{sec-extrap} are similar to those used in our ongoing studies of other $B$ and $B_s$ semileptonic decays.  In this section we discuss these efforts, including specific plans for each decay, and show preliminary results.

\subsection{$B_s\to K\ell \nu$ and $B_s \to \eta_s$}
$B_s\to K\ell\nu$ is expected to be measured by LHCb and/or BelleII and offers a prediction opportunity for lattice QCD.
We know of no existing lattice results but note another effort is currently underway~\cite{ASKtalk}.
Combined with experimental results, a lattice calculation of the form factors for this decay provides an alternative to $B\to\pi\ell\nu$ for the exclusive determination of $|V_{ub}|$.  The presence of a heavier spectator quark makes the lattice calculation simpler.

\begin{figure*}[t!]
\hspace{-0.07in}  
{\scalebox{0.98}{\includegraphics[angle=-90,width=0.53\textwidth]{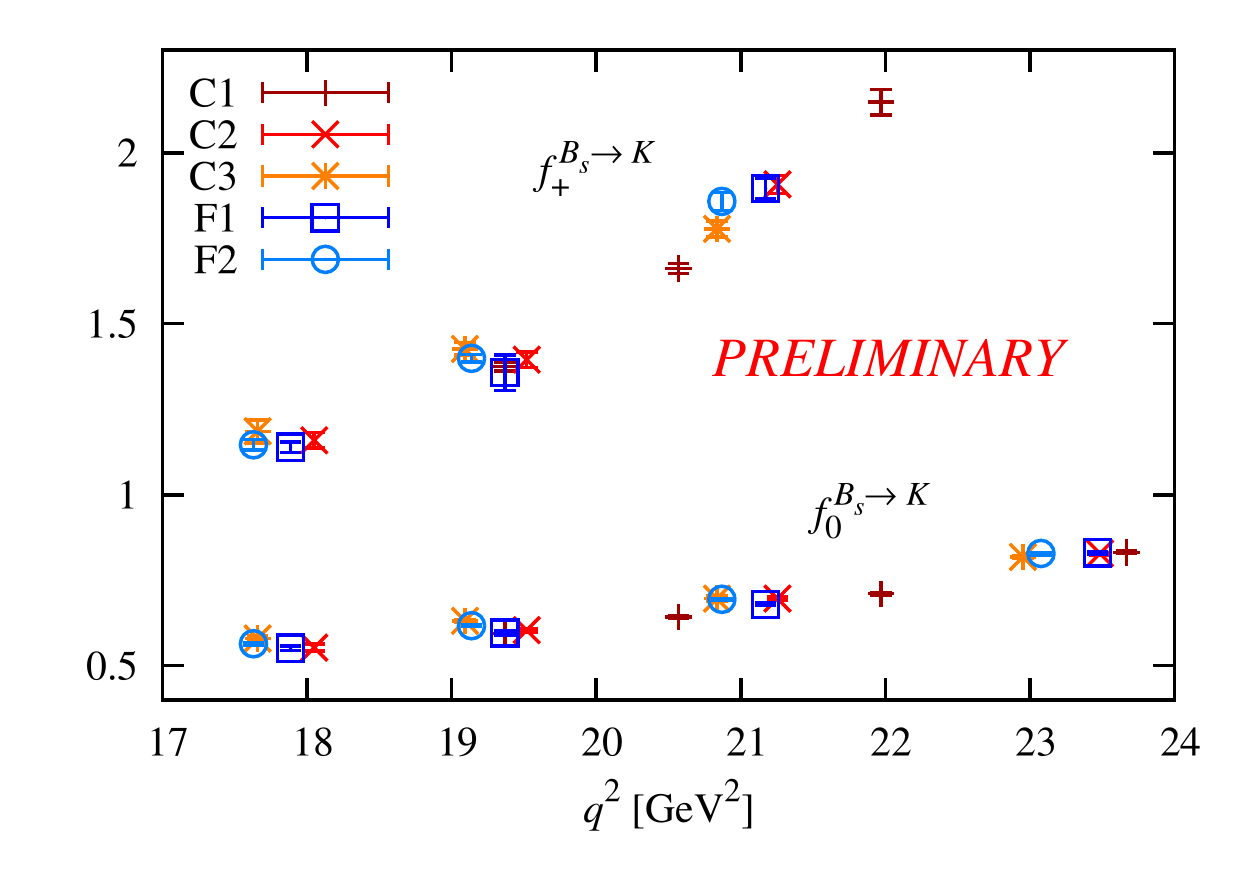}}}
{\scalebox{0.98}{\includegraphics[angle=-90,width=0.53\textwidth]{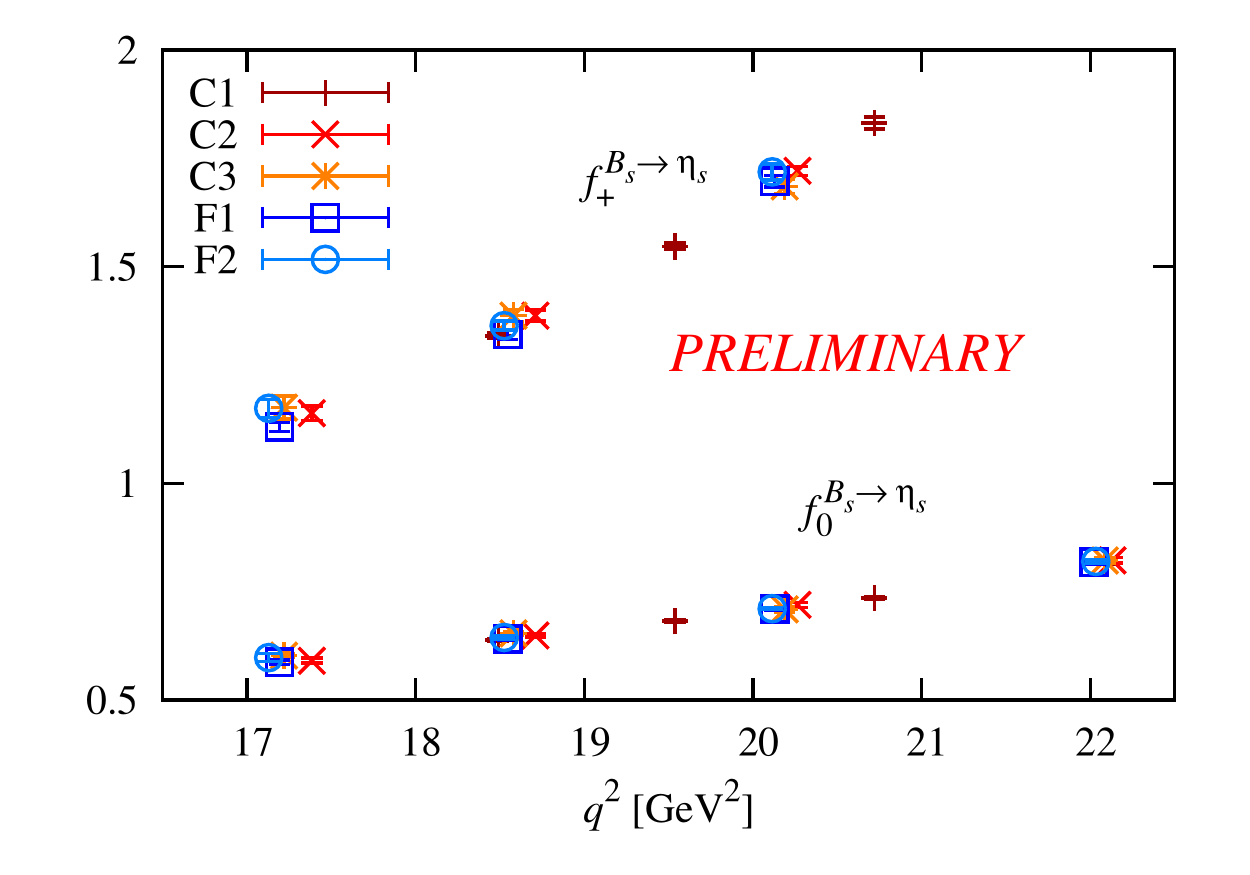}}}
\caption{Preliminary form factor results for (left) $B_s\to K\ell\nu$ and (right) $B_s\to\eta_s$.}
\label{fig-BsKEtas}
\end{figure*}

We plan to calculate ratios of form factors for $B_s\to K\ell\nu$ with those for the fictitious, lattice convenient process $B_s\to\eta_s$.  Combination with a future calculation of $B_s\to\eta_s$ using HISQ $b$ quarks should give $B_s\to K\ell\nu$ form factors with essentially no matching error, e.g.,
\begin{equation}
\left. \frac{ f_\parallel^{B_s\to K}(q^2_{\rm max}) }{ f_\parallel^{B_s\to \eta_s}(q^2_{\rm max}) } \right|_{{\rm NRQCD}\ b} \times\ \left. f_\parallel^{B_s\to \eta_s}(q^2_{\rm max}) \right|_{{\rm HISQ}\ b}.
\label{eq-ratio}
\end{equation}
More importantly, combining lattice values of $f_\parallel^{B_s\to K}(q^2_{\rm max})$ using NRQCD $b$ quarks will yield a nonperturbative determination of the matching factor for the $b\to u$ current, applicable to $B\to\pi\ell\nu$.
Preliminary results for $B_s\to K\ell\nu$ and $B_s\to\eta_s$ are shown in Fig.~\ref{fig-BsKEtas}.  To include correlations in the ratio of Eq.~(\ref{eq-ratio}) we are developing new fitting methods capable of simultaneously fitting large amounts of correlated data.

\subsection{$B\to D\ell\nu$ and $B_s\to D_s\ell\nu$}
\begin{figure*}[t!]
\hspace{-0.04in}  
{\scalebox{1}{\includegraphics[angle=-90,width=0.53\textwidth]{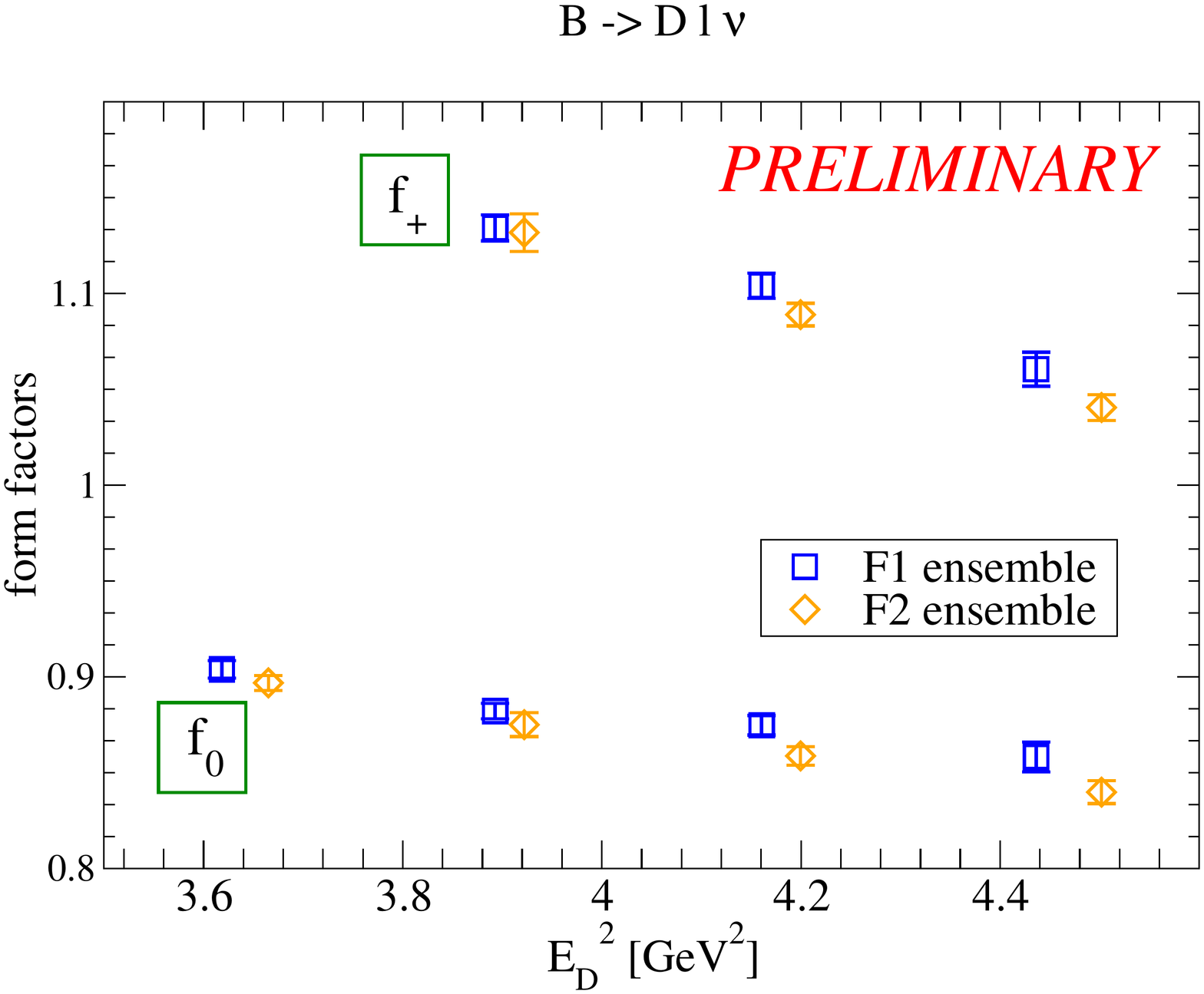}}}
{\scalebox{1}{\includegraphics[angle=-90,width=0.53\textwidth]{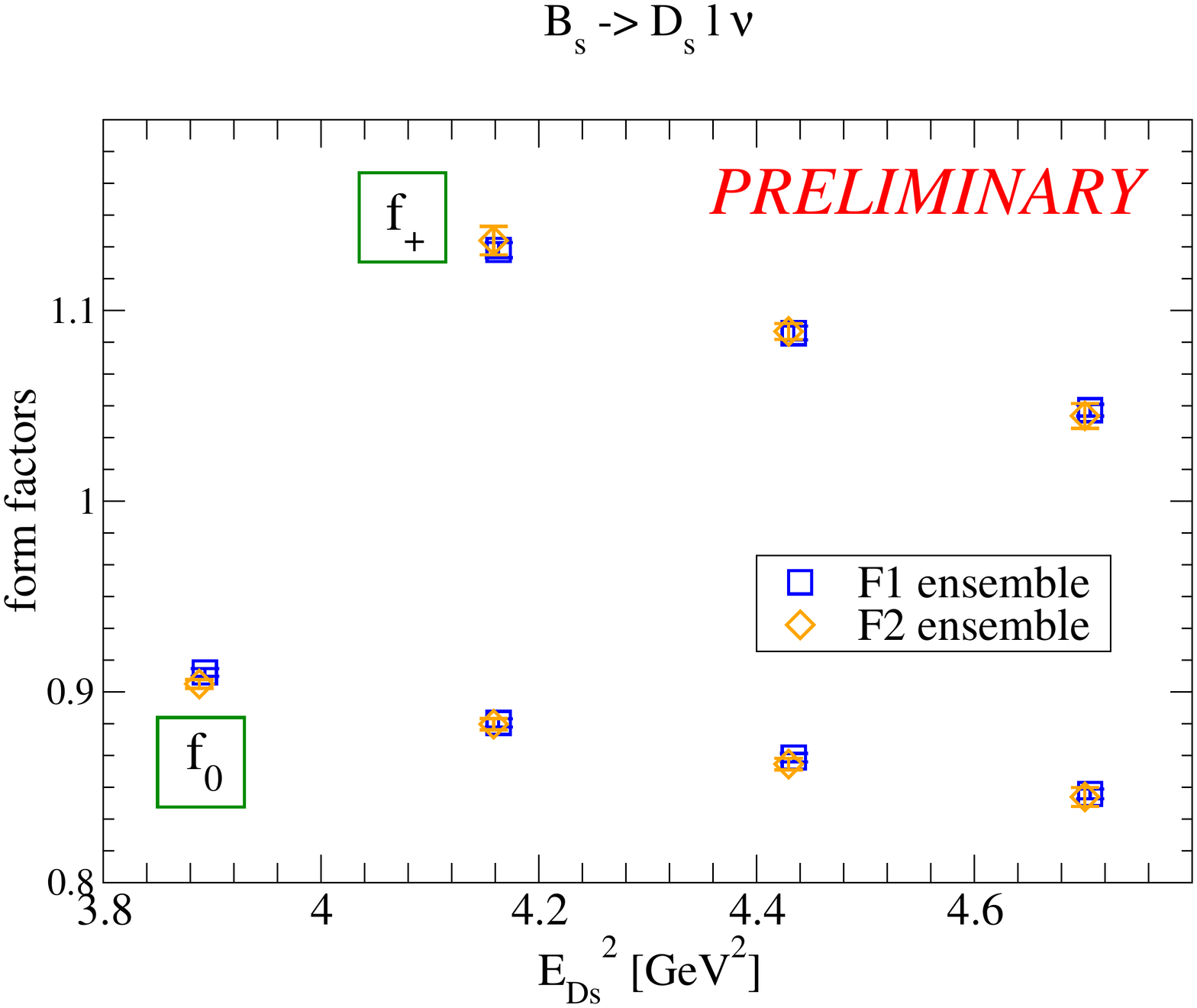}}}
\caption{Preliminary results for $B\to D\ell\nu$ and $B_s\to D_s \ell\nu$ form factors.}
\label{fig-BD}
\end{figure*}

These processes allow exclusive determinations of $|V_{cb}|$.
BelleII expectations~\cite{BelleIIproj} suggest they will not likely be competitive with the precision currently achievable in $B\to D^*\ell\nu$, but they would provide important input to help sort out the discrepancy between inclusive and exclusive determinations.
Furthermore, the ratios
\begin{equation}
\frac{ \mathcal{B}(B \to D\tau\nu) }{ \mathcal{B}(B \to D\ell\nu) } \hspace{0.2in} {\rm and} \hspace{0.2in} \frac{ f_0^{B_s\to D_s}(M_\pi^2) }{ f_0^{B\to D}(M_K^2) }
\end{equation}
provide a useful to constraint on new physics~\cite{FNALRD} and can be used in the determination of $\mathcal{B}(B_s\to \mu\mu)_{\rm expt.}$~\cite{FNALBs}, respectively.
Preliminary results on two fine ensembles are shown in Fig.~\ref{fig-BD}.
Other lattice works underway include~\cite{Qiutalk, Atouitalk}.  

\subsection{$B\to \pi\ell\nu$ and $B\to \pi\ell\ell$}
The SM tree-level decay $B\to\pi\ell\nu$ provides the ``standard" exclusive determination of $|V_{ub}|$.
The lattice contribution to $|V_{ub}|$ introduces an $(8-9)\%$ error~\cite{lattavg}, while experimental input has a $4\%$ error that is expected to halve in 5 years.
We are improving upon HPQCD's previous effort~\cite{oldBPi} by incorporating $b$ quark smearing, HISQ light valence quarks with random wall sources, improved scale determination and quark mass tuning, advances in correlator fitting methods, and the kinematic $z$ expansion.

The rare decay $B\to\pi\ell\ell$, which precedes via $b\to d$ FCNC, has been observed for the first time in~\cite{LHCbBPi} and future improvements are expected.
As with $B\to K\ell\ell$ we accommodate generic new physics by calculating the scalar, vector, and tensor form factors.
%
\begin{figure*}[t!]
\hspace{-0.07in}  
{\scalebox{0.98}{\includegraphics[angle=-90,width=0.53\textwidth]{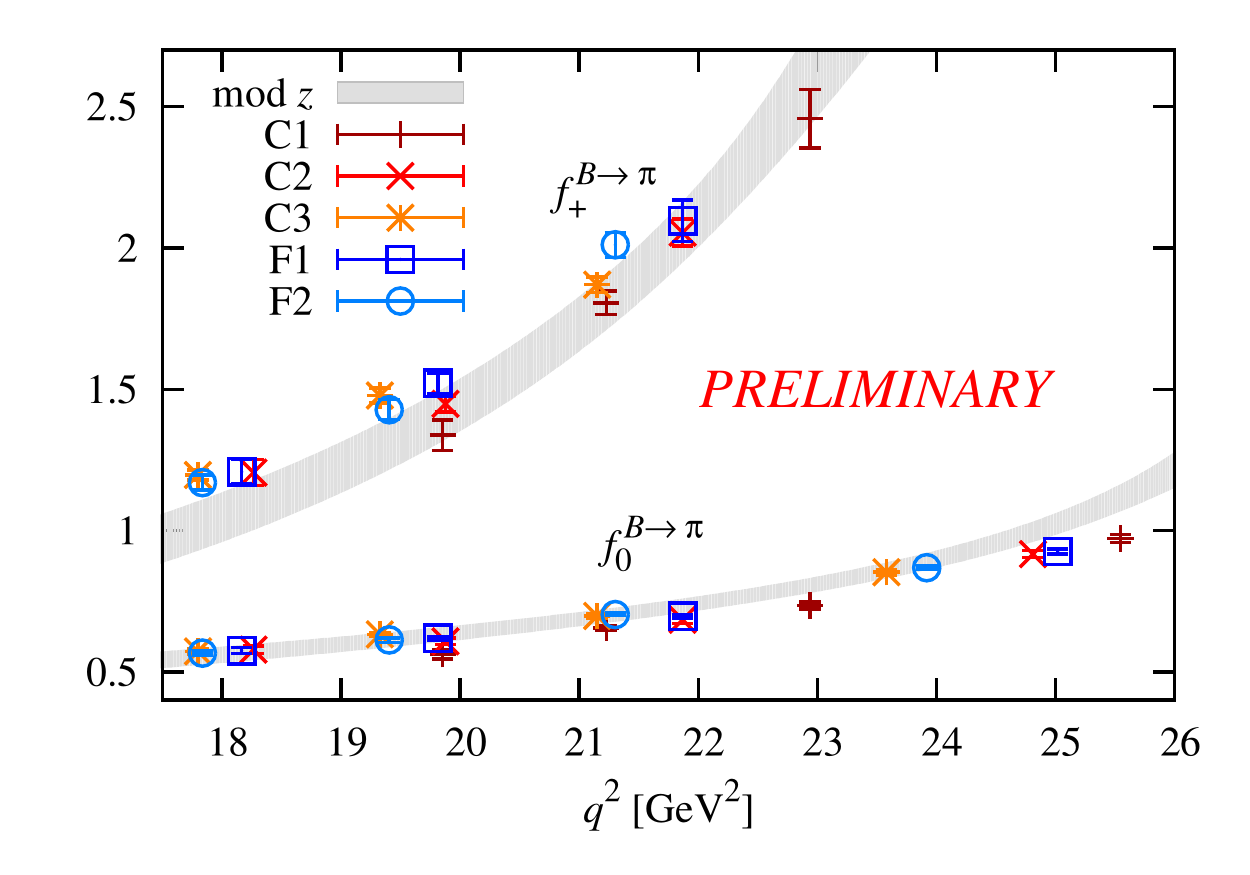}}}
{\scalebox{0.98}{\includegraphics[angle=-90,width=0.53\textwidth]{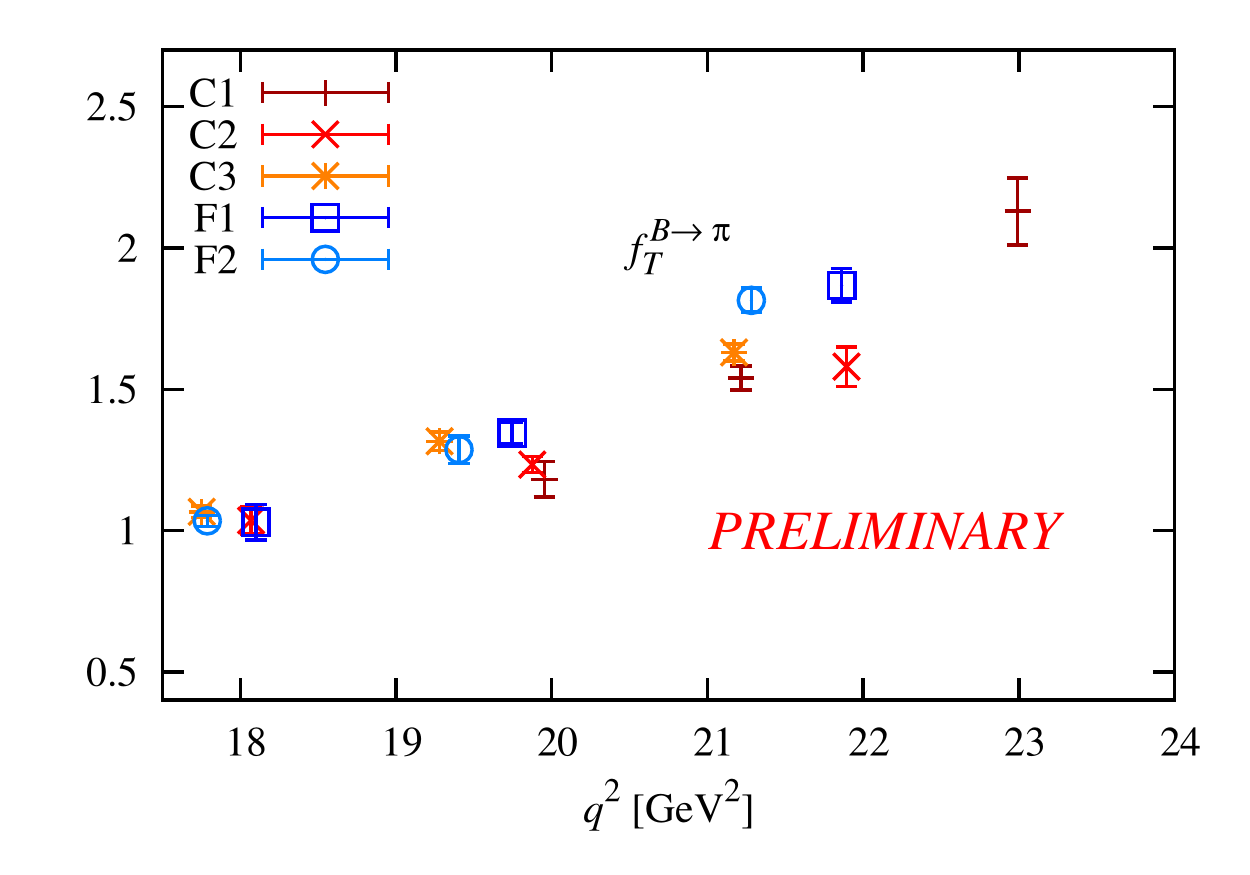}}}
\caption{Preliminary results for $B\to \pi\ell\nu$ and $B\to \pi \ell\ell$ form factors.}
\label{fig-BPi}
\end{figure*}
Fig.~\ref{fig-BPi} shows preliminary results, including a chiral-continuum and kinematic extrapolation via the modified $z$ expansion.  Based on these preliminary results we have increased statistics and added momentum $\nicefrac{L{\rm \bf p}_K}{2\pi} = 200$ for select ensembles.

\section{Next steps}
We are developing correlator fitting methods for simultaneous fits to large correlated data sets (ie. those involved in ratios).
To better accommodate simulation data at energies $\gtrsim 1$ GeV, we are considering hard pion chiral perturbation theory~\cite{HPChPT} and studying the effectiveness of the modified $z$ expansion.

\section*{Acknowledgements}
This research was supported by the DOE and NSF.  We thank the MILC collaboration for making their asqtad $N_f=2+1$ gauge field configurations available.  Computations were carried out at the Ohio Supercomputer Center and on facilities of the USQCD Collaboration funded by the Office of Science of the U.S. DOE.

\end{document}